\documentclass[aps,twocolumn,10pt]{revtex4}
\usepackage{amsfonts}
\usepackage{amsmath}
\usepackage{amssymb}
\usepackage{graphicx}
\usepackage{epsfig}

\begin{document}

\title{ Drag effect and topological complexes in strongly interacting
two-component lattice superfluids}
\author{V. M. Kaurov$^{a)}$, A. B. Kuklov$^{a)}$ and A. E. Meyerovich$^{b)}$}
\affiliation{$^{a)}$Department of Physics, CUNY - Staten Island, New York, NY 10314\\
$^{b)}$Department of Physics, University of Rhode Island, Kingston, RI 02881}

\begin{abstract}
The mutual drag in strongly interacting two-component superfluids in optical
lattices is discussed. Two competing drag mechanisms are the
vacancy-assisted motion and proximity to the quasi-molecular state, in which
an integer number $q$ of atoms (or holes) of one component might be bound to
one atom (or hole) of the other component. Then the lowest energy
topological excitation (vortex or persistent current) becomes a composite
object consisting of $q$ circulation quanta of one component and one
circulation of the other. In the SQUID-type geometry, the value of $q$ can
become fractional. These topological complexes can be detected by absorptive
imaging. We present both the mean field and Monte Carlo results. The drag
effects in optical lattices are drastically different from the Galilean
invariant Andreev-Bashkin effect in liquid helium.
\end{abstract}

\maketitle

Multi-component quantum mixtures in optical lattice (OL) are a source of new
and rich many-body physics. The one-component bosons in OL have been
exhaustively studied theoretically \cite{BOSE_theor} and experimentally \cite%
{BOSE_exp}, especially in the context of the quantum phase transition
between superfluid (SF) and Mott insulator. The study of multi-component
boson systems in OL has just began. Theoretical investigations predict
variety of new quantum phases with unusual properties \cite%
{Zhou,SCF,meyer2002,We1,We2,Demler}. Two interesting recent examples include
topological excitations -- vortices and persistent currents with
non-standard winding properties in two-component superfluids (2SF) \cite%
{meyer2002,We1}.

Crucial, but largely unaddressed effect is the impact of strong interaction
on properties of superfluid phases where each component $\psi _{a}$ has its
finite expectation value $\langle \psi _{a}\rangle $. %In a one-component
%superfluid, the strong interaction causes large depletion of the condensate.
%The same is expected in 2SF. Besides depletion, another 
Interesting manifestation of the strong interaction is the inter-component
drag similar to the Andreev-Bashkin effect \cite{Andreev} in 2SF helium
mixtures. In general, the drag effect between non-convertible species at
zero temperature is represented by the cross-terms in the expansion of the
ground state energy in terms of small gradients of the superfluid phases $%
\nabla \varphi _{a},\,a=1,2$, 
\begin{equation}
\delta E=\int d\mathbf{x}[\frac{1}{2}\sum_{a,b}\rho _{ab}\nabla \varphi
_{a}\nabla \varphi _{a}],  \label{twist}
\end{equation}%
with $\rho _{ab}$ standing for the superfluid stiffnesses. The cross-term $%
\rho _{12}$ is responsible for the drag. It is due to interaction effects and is not confined to some
particular term in the full microscopic many-body Hamiltonian. Depending on
its sign, this term describes either a mutual unidirectional flow or a
counterflow of the components. The Galilean invariance argument, often
attributed to Landau, imposes two constraints on $\rho _{ab}$. These
constraints are responsible for the Andreev-Bashkin effect in superfluid
mixtures of liquid helium isotopes in which $\rho _{12}$ is uniquely related
to the ratio of bare $m_{1}$ and effective $m_{1}^{\ast }$ masses of
minority atoms in the host superfluid of the majority component. The
Galilean transformation to a frame moving with velocity $\mathbf{V}$
requires that the phase of each component changes as $\varphi
_{i}\rightarrow \varphi _{i}-\left( m_{i}/\hbar \right) \mathbf{V\cdot r}$,
where $m_{i}$ are the \textit{bare} masses. The energy density (\ref{twist})
transforms as $\delta E\rightarrow \delta E-\mathbf{PV}$, where $\mathbf{%
P/\hbar }=N_{1}\mathbf{\nabla }\varphi _{1}+N_{2}\mathbf{\nabla }\varphi _{2}
$ is the momentum density expressed in terms of particle densities $N_{1,2}$
of each component. This yields 
\begin{equation}
\rho _{11}m_{1}+\rho _{12}m_{2}=N_{1},\ \rho _{12}m_{1}+\rho
_{22}m_{2}=N_{2}.  \label{const}
\end{equation}%
Introducing effective masses $m_{1}^{\ast },\,m_{2}^{\ast }$ as $\rho
_{11}=N_{1}/m_{1}^{\ast },\,\rho _{22}=N_{2}/m_{2}^{\ast }$, we reproduce
the result \cite{Andreev}, $\rho _{12}=(N_{1}/m_{2})(1-m_{1}/m_{1}^{\ast })$
as well as the relation $N_{1}(m_{1}-m_{1}^{\ast })m_{1}/m_{1}^{\ast
}=N_{2}(m_{2}-m_{2}^{\ast })m_{2}/m_{2}^{\ast }$. In other words,
conservation of the total momentum requires that the difference of the bare
and effective masses is compensated by the flux of the other component. Note
that $\rho _{12}>0$ since $m_{1,2}^{\ast }>m_{1,2}$.

In the case of strong mass renormalization, $\left( m_{1}^{\ast
}/m_{1}\right) -1\geq 1$, quite spectacular effects should be expected \cite%
{Meyerovich} from the topological excitations - vortices. Specifically, the 
\textit{lowest energy} single-circulation vortex of the majority component ($%
\rho _{22}\gg \rho _{11}$) should carry several circulation quanta $q=1,2,..$
of minority component. The equilibrium value of $q$ is obtained by
minimizing the factor $m_{2}q^{2}+2\left( m_{1}^{\ast }-m_{1}\right) q$ in
the energy of the vortex complex (or persistent current). These $q+1$ vortex
complexes exhibit transformations with respect to the value of $q$ depending
on external conditions that determine the value of $m_{1}^{\ast }$. If the
interaction is weak, $\rho _{12}$ can be calculated as an expansion in the
gas parameter \cite{Shevchenko}.

In this paper, we address the drag effect in a \textit{lattice} 2SF in
strongly interacting limit, and show that it is radically different from the
Galilean-invariant case. The lattice plays a central role in violating the
relation \cite{Andreev} between $\rho _{12}$ and $m_{1}/m_{1}^{\ast }$ (and
the constraints (\ref{const})). We also argue that the value of $q$ is
affected by proximity of the 2SF to the quasi-molecular phase.

In OL, in contrast to the Galilean-invariant system, the lattice provides a
preferred reference frame, so that the (hydrodynamic) properties of the
two-component mixture are determined not by the relative velocity of
components but by their individual velocities with respect to the lattice.
Furthermore, the effective mass in OL is formed largely by the width and
depth of laser-generated potential wells rather than by a trailing cloud of
the second component. Another crucial difference is that in OL number of
vacancies is a conserved quantity. Below we perform the mean field and Monte
Carlo analysis of the mutual drag in 2SF in three different physical
situations: a soft-core system close to molecular condensation, a hard core
system with finite intercomponent exchanges, and a hard
core system with vacancy-assisted motion without the intercomponent exchanges. 

\textit{Drag due to proximity to the quasi-molecular state}. Here we discuss
a generic mechanism leading to the $q+1$-topological complexes in the 2SF.
Strong drag effect occurs if a two-component boson system is close to a
transition into the quasi-molecular state in which the only broken symmetry
has the order parameter $\Phi _{q}\sim \exp (i\varphi ^{(q)})\sim \langle
\psi _{1}\psi _{2}^{q}\rangle \neq 0$ (or $\langle \psi _{1}\psi
_{2}^{\dagger q}\rangle \neq 0$). In pure molecular state with undefined
individual phases $\varphi _{1,2}$ (that is, $\langle \psi _{1,2}\rangle =0$%
), the phase-gradient energy is given by the molecular superfluid phase $\varphi
^{(q)}$ as $\delta E=\int d\mathbf{x}\rho _{q}(\nabla \varphi ^{(q)})^{2}/2$%
, with $\rho _{q}$ being molecular superfluid stiffness. The molecular order
parameter persists in the 2SF phase so that the additional broken $U(1)$
symmetry emerges continuously \cite{We1}. The two phases $\varphi _{1,2}$
become well defined in the 2SF state with the molecular phase being locked
as 
\begin{equation}
\varphi ^{(q)}=\varphi _{1}+q\varphi _{2}.  \label{mSF}
\end{equation}%
This locking can be understood as a consequence of virtual processes of
transformation of a $(q+1)$-molecule into $q$ B-atoms and one A-atom. The
corresponding contribution to the energy functional is %\begin{equation}
$\Delta E\sim \int d\mathbf{x}\Phi _{q}\psi _{1}^{\ast }\psi _{2}^{\ast
q}+H.c.$. %  \label{MF}
%$\end{equation}%
This term (\textit{cf.} the diatomic molecules with $q=1$ \cite{Tim})
ensures the relation (\ref{mSF}) in the longwave limit. Then the 
energy (\ref{twist}) becomes 
\begin{eqnarray}
\delta E &=& \int d\mathbf{x}[\frac{\rho _{q}}{2}(\nabla (\varphi
_{1}+q\varphi _{2}))^{2} \\
&+&\frac{\rho _{1}^{\prime }}{2}(\nabla \varphi _{1})^{2}+\frac{\rho
_{2}^{\prime }}{2}(\nabla \varphi _{2})^{2} + \rho_{12}^\prime \nabla
\varphi _{1} \nabla \varphi _{2}],  \notag  \label{m-twist}
\end{eqnarray}%
with $\rho _{ij}^{\prime }$ continuously changing from zero in the molecular
phase to some finite values in the 2SF phase. It is important that the
molecular stiffness $\rho _{q}$ is not a critical property of the system -
it does not change while crossing the phase boundary. Thus, at least close
to the phase boundary, minimization of the vortex energy gives $\varphi
_{1}=-q\varphi _{2}$, that is, the $q+1$ vortex. In reality, the relations $%
|\rho _{ab}^{\prime }|\ll |\rho _{12}|\approx \rho _{1,2}$ can hold quite
far from the phase boundary. This implies that the $q+1$ topological
excitation exist deep in the 2SF phase. We demonstrate this numerically for $%
q=1$ (see Fig.1 below).

It is convenient to introduce the drag coefficient $k$ as a ratio $k=\rho
_{12}/\rho _{11}$ of the cross-stiffness to the smallest diagonal stiffness, 
$\rho _{11}\leq \rho _{22}$. Then, as the minimization of the energy (\ref%
{twist}) shows, when $|k|>0.5$, a vortex of the dominant component can lower its energy
if it carries the circulation of the other component $q=\pm 1$.
In symmetric case ($\rho_{11}=\rho_{22}$), the integer $q$ closest to $k$ determines the $q+1$ vortex
(or persistent current) as the minimal topological excitation. It is
important to note that even small $|k|$ causes attraction between either
vortices of equal circulations ($k<0$) or between vortex and anti-vortex ($%
k>0$) in different components, so that if both exist they will form a
complex. Crossing the boundary $|k|>0.5$ has strong impact on mechanisms of
vortex creation and stability. For example, stirring the component with the
largest stiffness ($\rho _{22}$) above the threshold will cause creation of
the complex instead of a single vortex of the stirred component. Also, a
single vortex of the component $2$ becomes unstable with respect to inducing
creation of vortex of the other component.

The Hubbard lattice model with molecular phases,%
\begin{equation}
H=\sum_{\alpha ,<ij>}[-t_{\alpha }a_{\alpha ,i}^{\dagger }a_{\alpha
,j}+H.c.]+\sum_{\alpha ,\alpha ^{\prime },i}[U_{\alpha ,\alpha ^{\prime
}}n_{\alpha i}n_{\alpha ^{\prime },i}],  \label{soft}
\end{equation}%
has been extensively studied analytically \cite{SCF,Demler} and numerically 
\cite{We1,We2}. Here $U_{\alpha ,\alpha ^{\prime }}$ is the interactions
matrix, $t_{\alpha }$ describes the nearest-neighbor jumps of component $%
\alpha $; $a_{\alpha ,i}^{\dagger },\,a_{\alpha ,j}$ are the construction
bosonic operators, and $n_{\alpha i}=a_{\alpha ,i}^{\dagger }a_{\alpha ,i}$
are the on-site occupancies. As discussed in Refs.\cite{We1,We2}, the
quasi-molecular phase ($U_{12}<0$), namely, the paired superfluid, is in
many respects isomorphic to the super-counterfluid state ($U_{12}>0$) \cite%
{SCF} . Both states can undergo second order phase transition into the 2SF
phase so that the order parameter $\Phi _{q=1}=\langle \psi _{1}\,\psi
_{2}\rangle $ (or $\Phi _{q=1}=\langle \psi _{1}\,\psi _{2}^{\dagger }\rangle $%
) remains finite and robust. Obviously, in the 2SF phase, the $q=1$
composite vortices are the lowest topological excitations. As pointed out 
in ref.\cite{Mueller}, the Hamiltonian (%
\ref{soft}) also allows  molecular phases with arbitrary
integer value of $q$. This issue, though, requires separate analysis.

\textit{Hard core limit }$U_{ab}\rightarrow \infty $\textit{\ of the
Hamiltonian} (\ref{soft}). This limit can exhibit quite interesting physics
of strong quantum fluctuations even far from any phase transition \cite%
{meyer2002}. Obviously, when $N_{A}+N_{B}=1$ ($N_{A},\,N_{B}$ are the
average on-site occupancies of the species A, B), the system in the hard
core limit (HC) is the Mott insulator. Its ground state is degenerate with
respect to possible permutations of bosons A and B. This degeneracy, which
is a consequence of the HC approximation, is lifted by any infinitely small
inter-component exchanges. Accordingly, the two-component HC model should be
considered as a limit of the model in which the inter-component interaction $%
V_{int}=U_{12}$ is finite and increasing. In contrast to free space,
increasing $V_{int}$ leads to decrease of \textit{all} superfluid
stiffnesses because all transport is suppressed as $\sim t_1t_2/V_{int}$.
 Furthermore, in the limit $%
V_{int}\rightarrow \infty $ all stiffnesses are equal in magnitude. This is
clearly at variance with the free space constraints (\ref{const}) which
prohibit uniform decrease of all stiffnesses at fixed densities.

The two-component Hamiltonian with residual soft-core inter-component
repulsion is represented in terms of the HC construction operators $%
a_{i}^{\dagger },\,a_{i}$ and $b_{i}^{\dagger },\,b_{i}$ with Pauli
commutation relations for the A and B components 
\begin{equation}
H=\sum_{<ij>}[-t_{1}a_{i}^{\dagger }a_{j}-t_{2}b_{j}^{\dagger
}b_{i}+H.c.]+\sum_{i}V_{int}a_{i}^{\dagger }a_{i}b_{i}^{\dagger }b_{i}
\label{H_soft}
\end{equation}%
with summation $<ij>$ over the nearest-neighbor sites. At total filling 1,
this Hamiltonian has two phases - 2SF, where both SF order parameters are
defined, and super-counterfluid (SCF), where the only SF order is observed
in $\langle a_{i}b_{j}^{\dagger }\rangle $. 
%%%%%%%%%%%%%%%%%%%%%%%%%%%%%%%%%%%%%%%%%%%%%%%%%%%%%%%%%%%
%\begin{figure}[tbp]
%\includegraphics[bb=70 190 550 590, angle=-90, width=0.55\columnwidth]{fig1.ps}
\begin{figure}[h]
\begin{center}
\epsfxsize=8.5cm \epsfbox{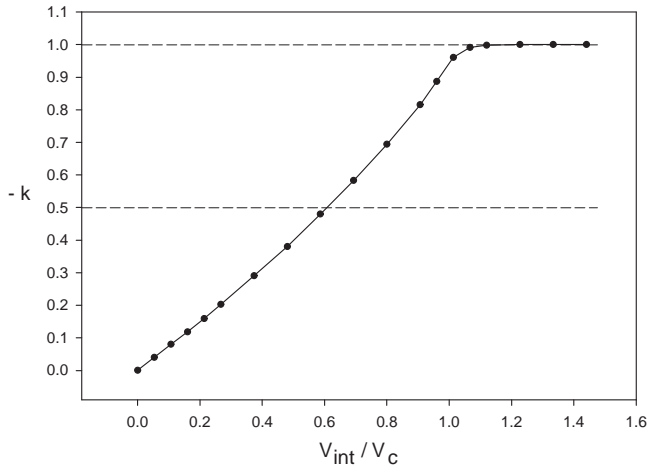}
\end{center}
\caption{ The drag coefficient $k$ for the J-current analog of the
Hamiltonian (\protect\ref{H_soft}) as a function of the relative
interaction, with the value $V_{int}/V_{c}=1$ corresponding to the 2SF-SCF
phase transition. The horizontal dashed lines indicate a domain where $1+q$%
-vortex complex with $q=1$ has lower energy than any single circulation
vortex. Error bars are much smaller then symbol sizes. Solid line is the eye
guide.}
\label{fig1}
\end{figure}
%%%%%%%%%%%%%%%%%%%%%%%%%%%%%%%%%%%%%%%%%%%%%%%%%%%%%%%%%%%
Transition between these two phases is continuous in the universality class
U(1) \cite{We1} and occurs in the symmetric case $t_{1}=t_{2}=t$ at some
value $V_{int}=V_{c}$, $V_{c}/t\sim 1$. As discussed above, the drag effect
is strong in the 2SF phase even far from the transition. We proved this by
performing the Worm algorithm \cite{worm} Monte Carlo simulations of the
two-color J-current model \cite{J-current,We1,We2} at zero temperature on a $%
2D$ square lattice. This model is a discrete-time grand-canonical analog of
the Hamiltonian (\ref{H_soft}) with the hard-core constraints. The
stiffnesses were determined from the statistics of the winding numbers
similarly to Refs. \cite{wind,We1, We2}. The SCF phase was identified by
observing $\rho _{11}=\rho _{22}=-\rho _{12}$. The negative value of $\rho
_{12}$ is due to counterflow of the components - each winding of A-worldline
is accompanied by opposite winding of B-worldline. In Fig.1, the drag
coefficient $k$ is plotted as a function of the relative interaction
strength. As can be seen, the domain $1/2<|k|<1$ in the 2SF (between the
dashed lines), where the composite $1+q$-vortex with $q=1$ has lower energy
than any single vortex, is not restricted to the vicinity of the critical
point $V_{c}$ but occupies about half of the phase diagram. Here $\rho
_{12}<0$, indicating that both components participate in the counterflow
even in the 2SF state.

\textit{Vacancy assisted drag}. If the total filling is different from 1,
system is always in 2SF phase at $T=0$. In this case, another mechanism
contributes to the drag -- the vacancy assisted transport. Atoms tunnel to
the \emph{unoccupied } sites (vacancies) much faster than the rate of the
A-B exchange with large $V_{int}$. The vacancies stimulate mass flow in one
direction and move in the opposite one. As a result, both components A and B
move in one direction, which means that $\rho _{12}>0$. This situation implies
crossover when $\rho _{12}$ changes sign at some special point with \textit{%
no drag,} $\rho _{12}=0$. Since no symmetry change takes place, this is not
a phase transition. The crossover from $k<0$ to $k>0$ takes place as $V_{int}
$ increases at fixed number of vacancies.

Note that the drag coefficient $k$ must increase when the number of
vacancies $x_{v}=1-N_{A}-N_{B}$ \textit{decreases}. This counterintuitive
result stems from the nature of vacancies. In one component case,
conservation of the number of vacancies $N_{V}$ makes them similar to
particles. The HC limit links the flow of vacancies with the opposite flow
of atoms. In the two-component case, the situation is similar with one
crucial difference -- a vacancy is not uniquely associated with a particular
sort of atoms. Thus, motion of a \textit{single} vacancy through a lattice in one
direction leads to flows of \textit{both} components in the opposite
direction. This implies strong drag with positive $k$. When $x_v$ increases,
system becomes more like a low density and, thus, weakly interacting mixture
of two sorts of atoms with correspondingly small $k$.

To analyze the mutual drag and the possibility of complex vortices in the
vacancy dominated regime we modified the HC model by imposing the additional
constraint $a_{i}b_{i}=0$ on (\ref{H_soft}) and introducing the chemical
potentials term $-\mu _{1}N_{A}-\mu _{2}N_{B}$ for each component to keep
control of the filling factors. As discussed in Ref.\cite{meyer2002}, this
limit can exhibit long range phase separation as well as short scale
fluctuative phase separation corresponding to minority particles acquiring
large cloud of vacancies.

If $1-N_{B}>N_{A}$, it is convenient to introduce a description in which the
vacuum corresponds to all sites filled by B particles. Then, the number $%
n=1-N_{B}$ of B holes is shared between $N_{A}$ atoms and remaining $%
x_{v}=n-N_{A}>0$ vacancies. In the limit $N_{A}\ll N_{B}\sim 1$ transport of
vacancies can be considered as transport of B holes with the effective
Hamiltonian 
\begin{equation}
H=\sum_{<ij>}[-t_{1}a_{i}^{\dagger }a_{j}v_{j}^{\dagger
}v_{i}-t_{2}v_{j}^{\dagger }v_{i}+H.c.],  \label{vac4}
\end{equation}%
where $v_{i}^{\dagger },\,v_{i}$ are the Pauli operators for B holes. In
order to describe the mutual drag within the mean field approximation, one
should replace the field operators $a,v$ by the functions $a=\sqrt{x_{1}}%
\exp \left( i\varphi _{1}\right) $, $v=\sqrt{x_{v}}\exp \left( -i\varphi
_{2}\right) $ with the slowly varying phases and perform the gradient
expansion. [The minus in front of $i\varphi _{2}$ indicates that flow of
holes and actual flow of mass are opposite]. This automatically generates
the term $\sim t_{1}x_{1}x_{v}(\nabla (\varphi _{1}+\varphi _{2}))^{2}$ in
effective energy from the first term in Hamiltonian (\ref{vac4}). Obviously,
the ratio of the stiffnesses becomes $k=\rho _{12}/\rho _{11}=1$ which
corresponds to \textit{positive} cross-term typical for the vacancy assisted
transport meaning that the mean field captures well the physics of the
vacancy assisted transport. However, the prediction $k=1$ and, therefore, $%
q=1$, is not supported numerically.

We have performed Worm algorithm \cite{worm} Monte Carlo simulations of the
two-color J-current model \cite{J-current,We1,We2} at zero temperature in $2D
$ square lattice in the HC limit with partial filling. 
%%%%%%%%%%%%%%%%%%%%%%%%%%%%%%%%%%%%%%%%%%%%%%%%%%%%%%%%%%%
%\begin{figure}[tbp]
%\includegraphics[bb=70 190 550 590, angle=-90, width=0.55\columnwidth]{fig2.ps}
\begin{figure}[h]
\begin{center}
\epsfxsize=8.5cm \epsfbox{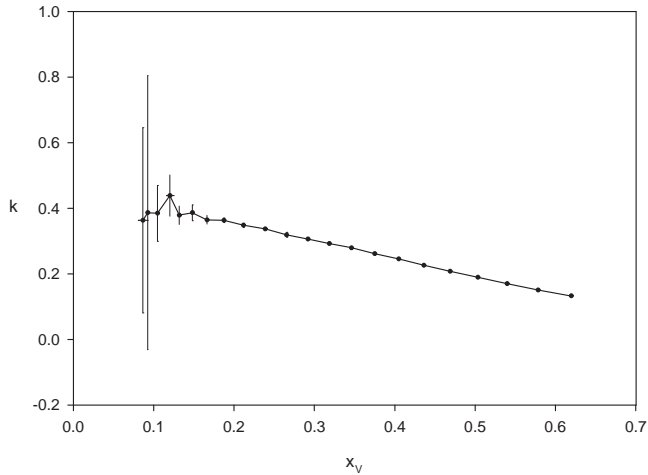}
\end{center}
\caption{ The drag coefficient $k$ for the J-model analog of the Hamiltonian
(\protect\ref{H_soft}) in the limit $V_{int}\rightarrow \infty $ as a
function of concentration of vacancies $x_{v}$ for symmetric case: $%
N_{A}=N_{B},\,t_{a}=t_{b}$. Error bars are shown for all points.Solid line
is the eye guide.}
\label{fig2}
\end{figure}
%%%%%%%%%%%%%%%%%%%%%%%%%%%%%%%%%%%%%%%%%%%%%%%%%%%%%
This model is similar to described above (\ref{H_soft}) with an additional
requirement of no double occupancy. We have found that (Fig.2), 
for $x_{v}\gtrsim 0.15$, $%
k<0.42\pm 0.02$, and, thus, no topological complexes
can exist as the lowest energy topological excitations in this regime. At
this point we do not have a simple explanation for this variance between
numerical and mean field results. Most likely, the mean field result is not
applicable for large $x_{v}$ in the symmetric mixture when the vacancies
cannot be uniquely identified with the holes in the majority component.

As the number of vacancies is tuned to become $x_{v}\leq 0.15$, all
stiffnesses exhibit large error bars which can be attributed to the regime
of strong quantum fluctuations \cite{meyer2002} associated with the
degeneracy of the ground state in the HC limit. The precise nature of this
effect requires separate analysis. For finite $V_{int}$, depending on $%
N_{A},\,N_{B}$, the ground state can exhibit various types of ordering
including the checkerboard insulator \cite{Demler}. Then, decreasing $x_{v}$
at $N_{A}=N_{B}\rightarrow 0.5$ will result in the first order phase
transition with strong fluctuations, similar to those in Fig.2, due to the
domain formation.

\textit{Fractional }$q$. In the case of finite drag with $|k|<0.5$ \textit{%
fractional} phase circulation $q=k$ can be observed when persistent current
is interrupted by a Josephson junction which lifts the requirement of the
integer of $2\pi $ windings by creating the phase jump across the junction.
Then, phase winding is determined solely by the minimization of energy.

\textit{Detection}. Finally, the $(q+1)$ -vortex complexes can be observed
by absorptive imaging technique similar to imaging of vortices in
one-component Bose-Einstein condensates \cite{Imaging}. Typical pattern
should include extra $q$ fringes in one component.

In summary, we explored generic mechanisms of drag effect in quantum bosonic
mixtures in optical lattice with hard and soft core interaction. Strong
mutual drag can result in composite topological structures. The drag in lattice
is not controlled by particle effective masses. The simplest mean field approximation
does not adequately describe the strong drag.

The authors acknowledge useful discussions with Erich Mueller, Nikolay
Prokof'ev and Boris Svistunov as well as the support by the grants NSF
PHY-0426814, PSC-CUNY (A.K.) and NSF DMR-0077266 (A.M.).

\end{document}